# Emergence of structural and dynamical properties of ecological mutualistic networks

Samir Suweis[1], Filippo Simini[2,3], Jayanth R. Banavar[4], Amos Maritan[1]

1 Dipartimento di Fisica 'G. Galilei' & CNISM, INFN, Università di Padova, Via Marzolo 8, 35131 Padova, Italy

2 Center for Complex Network Research and Department of Physics, Biology and Computer Science, Northeastern University, Boston, Massachusetts 02115, USA

3 Institute of Physics, Budapest University of Technology and Economics Budafokiut 8, Budapest, H-1111, Hungary

4 Department of Physics, University of Maryland, College Park, MD 20742, USA

**Mutualistic networks are formed when the interactions between two classes of species are mutually beneficial. They are important examples of cooperation shaped by evolution. Mutualism between animals and plants plays a key role in the organization of ecological communities[1-3]. Such networks in ecology have generically evolved a nested architecture[4,5] independent of species composition and latitude[6,7] - specialists interact with proper subsets of the nodes with whom generalists interact[1]. Despite sustained efforts[5,8,9,10] to explain observed network structure on the basis of community-level stability or persistence, such correlative studies have reached minimal consensus[11,12,13]. Here we demonstrate that nested interaction networks could emerge as a consequence of an optimization principle aimed at maximizing the species abundance in mutualistic communities. Using analytical and numerical approaches, we show that because of the mutualistic interactions, an increase in abundance of a given species results in a corresponding increase in the total number of individuals in the community, as also the nestedness of the interaction matrix. Indeed, the species abundances and the nestedness of the interaction matrix are correlated by an amount that depends on the strength of the mutualistic interactions. Nestedness and the observed spontaneous emergence of generalist and specialist species occur for several dynamical implementations of the variational principle under stationary conditions. Optimized networks, while remaining stable, tend to be less resilient than their counterparts with randomly assigned interactions. In particular, we analytically show that the abundance of the rarest species is directly linked to the resilience of the community. Our work provides a unifying framework for studying the emergent structural and dynamical properties of ecological mutualistic networks[2,5,10,14].**

Statistical analyses of empirical mutualistic networks indicate that a hierarchical nested structure is prevalent and is characterized by nestedness values that are consistently higher than those found in randomly assembled networks with the same number of species and interactions[1,6]. Nevertheless, the degree of nestedness varies among networks. Recently[5,10], it has been argued that nestedness increases biodiversity and begets stability, but these results are in conflict with robust theoretical evidences showing that ecological communities with nested interactions are inherently less stable than unstructured ones[12,14,15] and that mutualism could be detrimental to persistence[11,15]. We aim to elucidate general optimization mechanisms underlying network structure and its influence on community dynamics and stability.

There is a venerable history of the use of variational principles for understanding nature, which has led to significant advances in many sub-fields of physics, including classical mechanics, electromagnetism, relativity, and quantum mechanics. Our goal is to determine the appropriate variational principle that characterizes a mutualistic community in the absence of detailed knowledge of the nature and strengths of the interactions between species and their environment. We begin by showing that increases in the abundances of the species lead to an increase in the total number of individuals (henceforth referred to as the total population) within the mutualistic community. We then show that, under stationary conditions, the total population is directly correlated with nestedness and vice-versa. Finally, we demonstrate that nested mutualistic communities are less resilient than communities in which species interact randomly. These results suggest a simple and general optimization principle: key aspects of mutualistic network structure and its dynamical properties could emerge as a consequence of the maximization of the species abundance in the mutualistic community (see Figure 1).

We consider a community comprising a total of $S$ interacting species (see Methods Summary), in which population dynamics is driven by interspecific interactions. We model mutualistic and competitive species interactions using both the classical Holling type I and II functional responses[16,17,18] (Supplementary Information). We perform a controlled numerical experiment at the stable stationary state by holding fixed the number of species, the strengths of the interactions, and the connectance, and seek to maximize individual species population abundances by varying the network architecture. The simplest approach consists of repeatedly rewiring the interactions of a randomly drawn species so as to increase its abundance, i.e., each selected species attempts to change its mutualistic partners in order to enhance the benefit obtained from its interactions (see Methods Summary and Supplementary Information). The optimization principle may then be interpreted within an adaptive evolutionary framework within which species maximize the efficiency of resource utilization[19,20] and minimize their chances of becoming extinct due to stochastic perturbations[21,22]. Interestingly, we find that enhancements in the abundance of any given species most often results in growth of the total population along with a concomitant increase of the nestedness (see Figure 1). These results demonstrate the existence of a correlation between nestedness and species abundance and highlight a non-trivial collective effect through which each successful switch affects the abundances of all species, leading to an inexorable increase, on average, of the total number of individuals in the community.

In order to make analytic progress and to better understand the correlation found between the optimization of individual species abundances, the total number of individuals in the community and nestedness, we turn to a mean field approximation[5] in which the mutualistic (and competitive) interactions are assumed to have the same magnitude, $\sigma_\Gamma$ (and $\sigma_\Omega$). Within this approximation, we are able to prove that (see Supplementary Information for mathematical details): a) An increase in the abundance of any species more often than not leads to a net increase in the total population of the community; and b) Communities with larger total population have interaction matrices with higher nestedness and vice-versa. The intra specific (plant-plant and animal-animal) interactions play a key but secondary role compared to the mutualistic (plant-animal) interactions. The main effect of the intra specific interactions is to break the degeneracy in the network overlap (Supplementary Information). Extensive numerical simulations in the more general, non-mean field case of heterogeneous interactions also confirm these findings. The nestedness distributions of the optimized mutualistic networks shifts markedly to higher values than their random network counterparts (see Figures 2A-B and 3). Monte Carlo simulations substantiate the strong correlation between the total population and the nestedness of the mutualistic interaction network (Figure 2C). Our analytic calculations show that, for identical increments in population abundances, a community characterized by weak mutualism has a larger increase in nestedness than one with strong mutualistic interactions (Supplementary Information). This result suggests that, when the mutualistic interactions are strong, the network architecture may play a less crucial role than in the regime of weak mutualistic interactions, wherein optimal tuning of the architecture could lead to significant beneficial effects for the community.

Our results are very robust and do not depend on the details of the optimization algorithm, the initial condition or the transient dynamics. In addition to simulations of mutualistic communities starting with random interactions networks and then "reorganizing" toward a more optimal state, we also implement a more realistic scenario in which mutualistic communities progressively assemble and are optimized over the course of evolutionary timescales[23,24]. Indeed, we find that the final result is the same, i.e., the final optimized networks display a nested architecture (Supplementary Information). Remarkably, nested architectures in mutualistic communities could emerge from different initial conditions as a result of a rewiring of the interactions according to a variational principle aimed at maximizing either the fitness[25] of the individual insect/plant - whose surrogate is its species abundance - involved in the interaction swap (species level optimization) or the fitness of the whole community, measured by the total population of all species (community level optimization). The intriguing fact that these two optimization schemes lead to similar conclusions suggest that group selection mechanisms[26] may have played an important role in the evolution of cooperation among plant and pollinators[24].

Community persistence and stability are important dynamical properties characterizing ecological networks, but the way in which the two are related in real systems is far from trivial[21,22]. It has been suggested that mutualistic network structures lead to high community persistence[5,10]. Persistence, however, is only defined for systems out of their steady state[7,21], and is sensitive to initial conditions, transients dynamics, and to the system's distance from stationarity[11,13]. Here we focus on the study of community stability for the optimized stationary networks. Using perturbative expansion techniques, we analytically find that the abundance of the rarest species controls the stability of mutualisitic communities (see Supplementary Information for mathematical details). Moreover, the optimized networks result in spontaneous symmetry breaking with more abundant generalist species and less abundant specialist species (Figure 4A). The relatively low abundances of the specialist species make them more vulnerable to extinction and results in correspondingly lower community resilience, as measured by the maximum real part of the eigenvalues of the community matrix community (Figure 4B). The advantage of having a high total population leading to increased robustness against extinction due to demographic fluctuations, carries with it the cost associated with a lower resilience — the optimized network recovers from perturbations on a longer timescale than its random counterpart[12,14,27] (Figure 4C).

Several ecological factors, as well as evolutionary history, contribute to shaping empirical networks. Here we have shown how binary nested network architecture could emerge as a consequence of an optimization process or variational principle. An interesting unexplored issue is an analysis of emergent quantitative nestedness[12], i.e. the organization of the interaction intensities in the optimized networks along with a comparison to empirical data in mutualistic networks[12]. The framework proposed here ought to be applicable for investigating the possible driving forces sculpting mutualistic network architectures in a variety of systems ranging from social[28] to economic[29] and other biological[30] (e.g. protein interaction) networks.

## Methods Summary

**Interaction networks.** The initial interaction matrix $M$ is composed of four blocks: two diagonal blocks describing direct competition among plants ($\Omega_{PP}$) and pollinators ($\Omega_{AA}$) and two off-diagonal blocks characterizing the mutualistic interactions between $n_P$ plants and $n_A$ pollinators ($\Gamma_{PA}$), and vice versa ($\Gamma_{AP}$). The total possible number of mutualistic interactions in each of these two latter blocks is equal to $n_A \times n_P$. The connectance $C_\Gamma$ ($C_\Omega$) represents the fraction of the mutualistic (competitive) interactions that are non-zero. Mutualistic interaction intensities $\gamma_{ij}^{AP}$ and $\gamma_{ij}^{PA}$ represent the increase of the growth rate of animal (plant) species $i$ per unit of plant (animal) biomass $j$ and they are assigned randomly from the distribution $|N(0,\sigma_\Gamma)|$, whereas the competitive interactions are distributed as $-|N(0,\sigma_\Omega)|$. Here, $N(\mu,\sigma)$ is the normal distribution with the mean $\mu$ and variance $\sigma^2$ chosen to have stability of the underlying population dynamics, i.e., $\sigma < \sigma_c$, where $\sigma_c$ is the critical strength threshold above which, with high probability, no stable fixed point dynamics exist[14].

**Optimization algorithm.** We start with an existing network; select a species, $j$, randomly and an existing link to one of its partner species $k$; we attempt a rewire between the $j$-$k$ and the $j$-$m$ links where $m$ is a potential alternative mutualistic partner, that is $\gamma_{jk}$ is interchanged with $\gamma_{jm}$. If the $j$-$m$ link already exists, i.e. $\gamma_{jm}$ is different from zero, the switch leads to an interchange of interaction strengths; otherwise the swap corresponds to rewiring the $j$-$k$ link to $j$-$m$. The switch is accepted if and only if it does not lead to a decrease of the population abundance of species $j$ in the steady state of the new network configuration. See

Supplementary Information for details.

**Supplementary Information** is linked to the online version of the paper at www.nature.com/nature.

**Acknowledgements** AM and SS acknowledge *Cariparo* foundation for financial support and *Studio 7 a.m.* for graphics support. We thank Stefano Allesina, Todd Cooke, Jacopo Grilli and Lorenzo Mari for useful discussions and three anonymous referees for invaluable comments and suggestions.

**Author Contributions**  SS carried out the numerical calculations and the data analysis. All the authors contributed to other aspects of the paper.

**Author Information** Author Information Reprints and permissions information is available at www.nature.com/reprints. The authors declare no competing financial interests. Readers are welcome to comment on the online version of this article at www.nature.com/nature. Correspondence and requests for materials should be addressed to SS (suweis@pd.infn.it) or AM (amos.maritan@pd.infn.it).

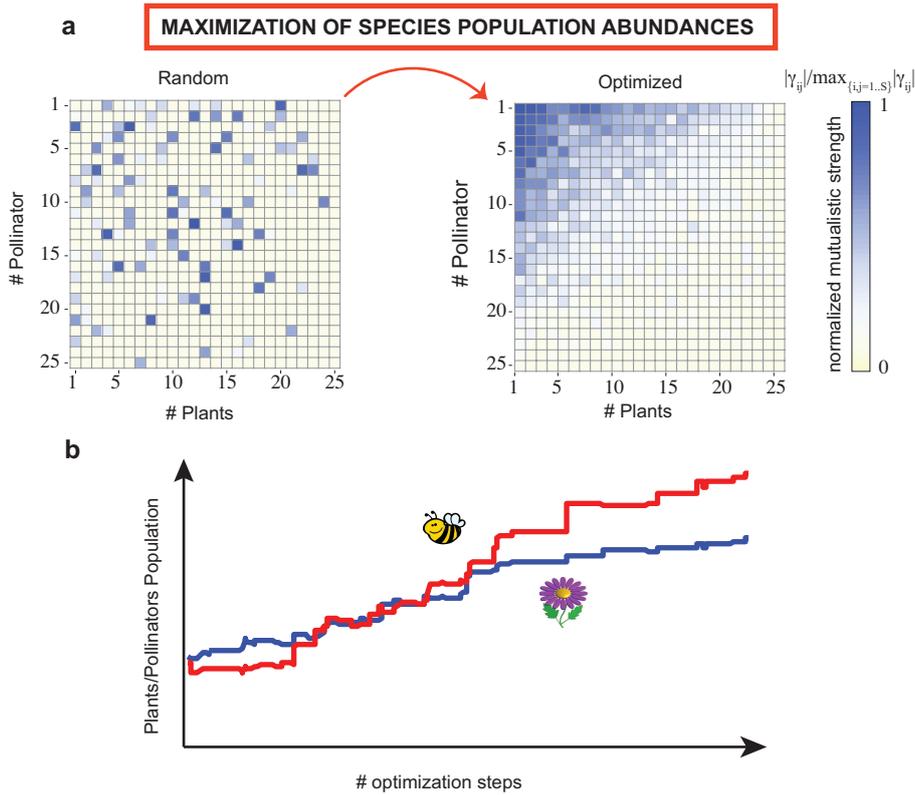

**Figure 1|The Optimization Principle** (a) The maximization of species abundance leads to networks with a nested architecture. The optimal interaction matrix shown is the typical architecture resulting from averages performed over 100 optimal networks starting from random realizations. The blue scale is a measure of the average mutualistic strength normalized with respect to the maximum strength interaction. (b) Because of mutualism, the optimization of the abundances of the individual species involved in the interaction rewiring results in an overall increase of the total population of both pollinators (in red) and plants (in blue). The curves represent the result of a typical run (no average is involved). Simulations presented here are obtained with Holling Type II dynamics and parameters $S=50$, $C_\Omega = C_\Gamma =4/S^{0.8}$ and $\sigma_\Omega=\sigma_\Gamma<\sigma_c$ (see Methods Summary).

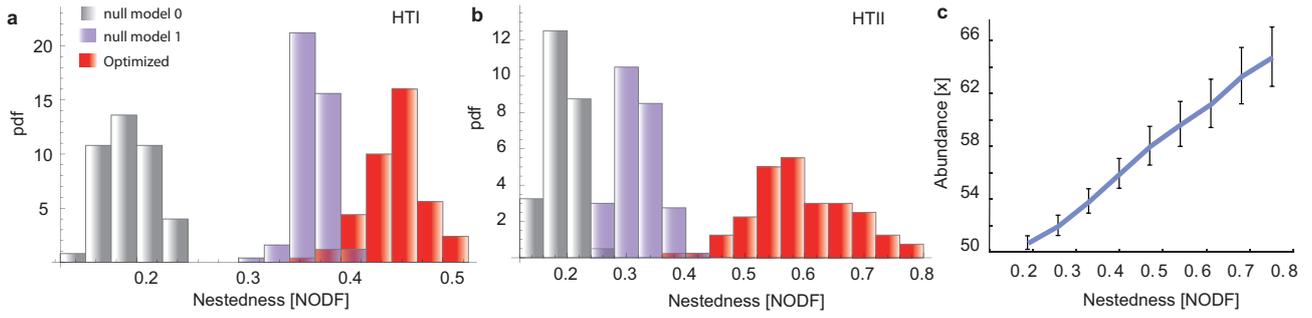

**Figure 2|Relation between nestedness and species abundance.** Histograms of the nestedness probability density for optimized mutualistic networks (shown in red) using the individual species optimization algorithm employing (a) Holling type I (*HTI*) and (b) type II (*HTII*) saturating functions. The histograms for the corresponding null model randomizations are also shown. In null model $0^{6,27}$, we preserve the dimensions and the connectivity of the optimized interaction network *M* with a random placement of the edges. In null model $1^{6,27}$, we also conserve the average number of connections for each plant and insect. The plots are obtained using 100 realizations of the optimization algorithm presented in the main text. In each realization, a new initial interaction matrix, *M*, is extracted with the same average $\mu=0$, variance $\sigma_\Omega=\sigma_\Gamma<\sigma_c$, and connectance $C_\Omega = C_\Gamma$ (see Methods Summary). (c) We consider interaction matrices ($S=50$, $C_\Omega=C_\Gamma$ and $\sigma_\Omega=\sigma_\Gamma<\sigma_c$) with different values of nestedness and we calculate the stationary population associated with each one of them: the nestedness and the total abundance of individuals in the community are strongly correlated. The connectance has been chosen to vary with the number of species as $C_\Gamma = 4S^{-0.8}$, obtained as a best fit to empirical data (Supplementary Information). Similar results are obtained for different parameter values and implementations of the optimization algorithm.

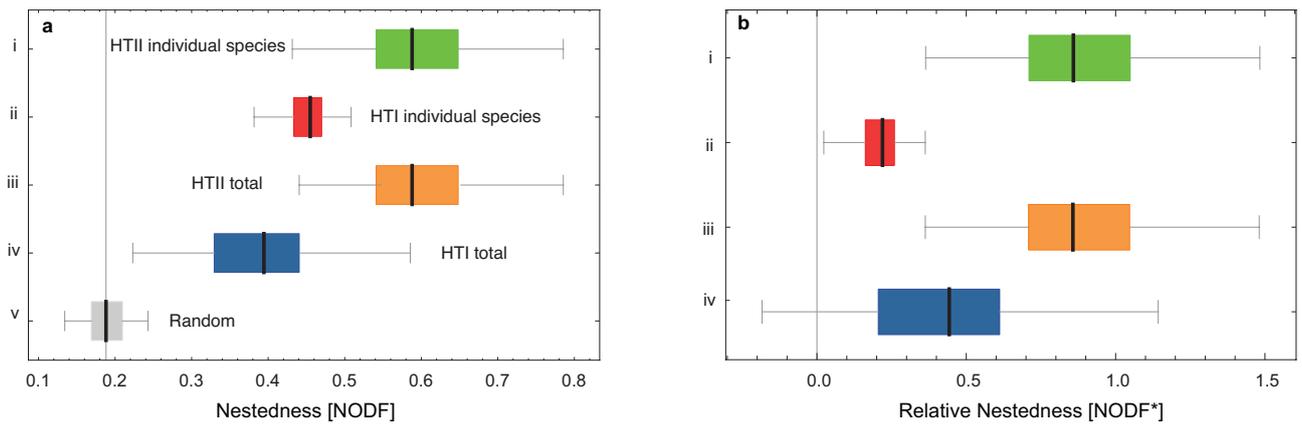

**Figure 3|Box-Whisker plots of the degrees of nestedness for optimized networks.** The ends of the whiskers represent the minimum and maximum, while the ends of the box are the first and third quartiles and the black bar denotes the median. The plots show (a) the absolute nestedness (NODF) and (b) relative nestedness[10] normalized to null model $1^{6,27}$ of 100 bipartite networks resulting from: species-level optimization (i-ii), community-level optimization (iii-iv) and random mutualistic networks - null model $0^{6,27}$ (v). Parameters used here are $S=50$, $C=4/S^{0.8}$ and $\sigma_\Omega=\sigma_\Gamma<\sigma_c$.

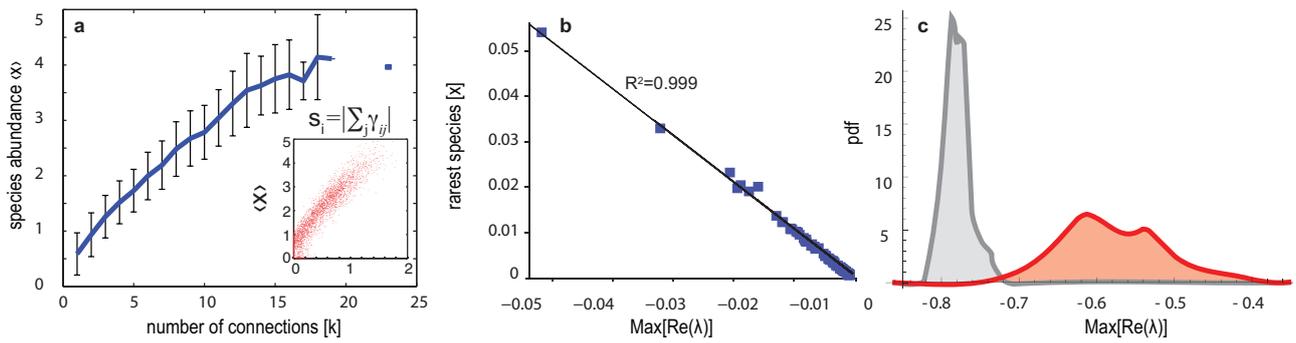

**Figure 4|Optimized networks are less resilient.** a) Average species abundance <x> as a function of the number of mutualistic partners of a species. The error bars represent the ±1 standard deviation confidence interval. Generalist species with more connections are, on average, more abundant than specialist species with fewer connections The red points in the inset depict <x> as a function of the mutualistic strength s. (b) Relationship between the abundance of the rarest species and system resilience given by the largest among the real parts (closest to zero) of the eigenvalues of the linearized stability matrix. The gray line shows a linear fit ($R^2$=0.999). (c) Probability density function (pdf) of the largest among the real parts of the eigenvalues - Max[Re($\lambda$)] - of the optimized community stability matrix (red curve) and of the corresponding initial random networks (gray curve). Optimized networks are less resilient than their random counterparts. The plots are obtained from 100 realizations of the community-optimization algorithm performed with HTI saturating function $S$=50, $C$=4/$S^{0.8}$ and $\sigma_\Omega=\sigma_\Gamma<\sigma_c$.